\documentstyle{article}

\begin{document}
\centerline{\bf THE SECRET LIFE OF THE DIPOLE}
\vskip .5cm
\centerline{Jeeva S. Anandan}
\vskip .5cm
\centerline{Department of Physics and Astronomy}
\centerline{University of
South
Carolina}
\centerline{Columbia, SC 29208, USA.}
\centerline{E-mail: jeeva@sc.edu}
\centerline{and}
\centerline{Department of Theoretical Physics}
\centerline{University of Oxford}
\centerline{1 Keble Road, Oxford OX1 3NP, U.K.}
\begin{abstract}

A new force on the magnetic dipole, which exists in the presence of both
electric and magnetic fields, is described. Its origin due to the `hidden 
momentum', implications and possible experimental tests are discussed.

\end{abstract}

We are familiar with the acceleration of charged particles, such as
protons and electrons, by uniform electromagnetic fields. But neutral
particles can be accelerated by uniform fields too, if they possess a
magnetic dipole$^{1,2}$. Some consequences and the possibility of
observing
this force are the focus of a new study$^3$.

The force on a magnetic dipole in an electromagnetic field has been
studied for more than 100 years. The expression found in textbooks depends
on the spatial gradient of the applied magnetic field: ${\bf F} = \nabla
({\bf \mu}\cdot{\bf B})$ where $\bf \mu$ is the magnetic moment and $\bf
B$ is the magnetic field. If we imagine the magnetic dipole to be an
infinitesimal current loop, then $\bf\mu$ is proportional to the angular
momentum or the spin of the circulating current. Then $\bf F$ may be
understood as the net force resulting from the different Lorentz forces
acting on different parts of the current loop due to the variation of the
magnetic field over the loop. 

It therefore came as a surprise to me to
find, nine years ago, that there is an additional force on the dipole
given by$^{1,2}$ \begin{equation} {\bf f} = \tau ({\bf B}'\times
{\bf\mu})\times {\bf E} \end{equation} where $\bf E$ is the electric
field, ${\bf B}'= {\bf B}-{\bf v}\times {\bf E}$ is the magnetic field in
the rest frame of the dipole that is moving with velocity $\bf v$ relative
to the laboratory, and $\tau$ is the ratio of $\bf \mu$ to the spin $\bf
S$ of the dipole (using units in which the velocity of light is 1).

This force was surprising because it exists even when the 
fields do not vary in position. It cannot therefore be obtained from 
the above intuitive picture. It was surprising also because of its 
dependence on the electric field: as the current loop 
representing the dipole may have no 
net electric charge it may be expected not to couple to 
an electric field. Also, (1) is doubly non linear; it does not reverse 
in direction if either the electromagnetic field strength or the 
magnetic moment is reversed (without reversing the spin). I know no other
force that is non linear in the field strength (as opposed to the 
potential). But after four separate derivations$^{1,2}$ of (1), in quantum
and 
classical physics, I was convinced that it does exist. 

A key to understanding the new force is that for this dipole the kinetic 
momentum (classically $m{\bf v}$) differs from the canonical 
momentum $\bf p$. The canonical momentum is conserved if there is
translational symmetry in space, as when the dipole interacts with a
uniform electromagnetic field. In fact,
\begin{equation}
m{\bf v} ={\bf p} -{\bf \mu}\times {\bf E}
\end{equation}
The rate of change of (2) is the total force by Newton's second 
law. The time derivative of $\bf p$ is $\bf F$, while the time 
derivative of the last term
is 
$\bf f$ (assuming that $\bf E$ is time independent). 

The last term in (2), which is due to relativistic effects, is called the
{\it hidden momentum}. This explains the title of this article.
The hidden 
momentum is like the potential for the dipole interacting with an 
appropriate $SU(2)$ Yang-Mills gauge field, because $\bf \mu$ is 
proportional to the spin vector whose 
components generate the $SU(2)$ group. Then (2) corresponds to 
the non linear term in the Yang-Mills field strength which 
determines the force.

An example of a neutral particle with a magnetic dipole is the neutron. As
in the above example of a current loop, the magnetic moment of the neutron
is proportional to its spin of magnitude $s=\hbar /2$, where $\hbar$ is
Planck's constant divided by $2\pi$. Hence, $\tau =\mu/s =2\mu /\hbar$. It
turns out that when $ B= 1$ Tesla and $E = 10^7~V~m^{-1}$, which can be
realistically achieved in the laboratory, the acceleration caused by the
new force (1) is of the order of $12$ cm/sec$^2$. On the face of it this
is a large acceleration. But the problem in actually detecting it is that
the spin precesses about ${\bf B}'$, due to the torque it experiences.
With a precession frequency of $1.8\times 10^8$ radians $s^{-1}$, $\bf f$
averages to zero very quickly. That is why it has not already been
observed in the large number of experiments in which a dipole interacts
with an electromagnetic field. I suggested that to make the effect of the
force accumulate, instead of cancel out, $\bf B$ may be kept constant
while $\bf E$ alternates in space$^{1,2}$.

Recently, in a beautiful and clearly written paper, Wagh and Rakhecha$^3$
have carefully studied this proposal in detail.
The neutron is assumed to pass through a sequence of cells
each having length $l$ and containing transverse uniform electric fields
$\bf -E, E, -E, E$.... such that $\bf E$ is parallel to a uniform
magnetic
field $\bf B$. The length $l$ is
chosen so that the time taken for the neutron to travel each cell is half
the Larmor period $T$ - that is, the time for its spin to rotate around
the
magnetic field ${\bf B}' $. Suppose that Cartesian axes are chosen such
that the $x-$ axis is along the direction of motion of the neutron and the
$z-$ axis is along the common direction of $\bf E$ and $\bf B$.  To lowest
order, ${\bf B}'={\bf B}$. Then (1) in this approximation is
\begin{equation} {\bf f} = {4\mu^2\over \hbar^2}BE(S_x ,S_y , 0)
\end{equation}

If originally $\bf S$ was in the $y-$ direction then in this approximation
it will rotate in the $x-y$ plane with constant frequency $2\pi /T$. As
the neutron traverses the first cell, $S_x$ is negative while $S_y$ is
both positive and negative for sucessive equal durations and averages to
zero. Since $E$ is negative in the first cell, $f_x$ is therefore
positive. In the next cell, $S_x$ is positive while $S_y$ is negative and
positive for successive equal durations. But owing to the reversal of $\bf
E$ in the second cell, $f_x$ is positive in the second cell also. Hence
$mv_x$ steadily increases, whereas $mv_y$ fluctuates.  
More exactly, because $\bf S$ rotates about $\bf B'$ which slightly tilts
from the $z-$ axis to its two sides in the $y-z$ plane alternatively in
successive cells, $\bf S$ actually spirals towards the $z-$ axis. The
spin $\bf S$
points in the $z-$ direction after the neutron traverses through $2n$
cells, where $n$ is a suitably chosen positive integer. Therefore, $f_x$
and $f_y$ gradually decrease, according to (3), which is reflected in a
corresponding tapering off of the changes in $v_x$ and $v_y$.

In the absence of the new force $\bf f$, the change in velocity is
entirely
due to $\bf F$ with $\bf B'$ substituted for $\bf B$. As the neutron
enters each cell, owing to the change in $\bf E$ and therefore ${\bf B}'$,
there would be a sudden change in its velocity. 
Only the $x-$ component of the velocity changes due to $\bf F$
because
of the translational symmetry in the $y-$ and $z-$ directions.  
This component varies as a step function of the number of cells traversed.
(The
splitting of the wave packets due to the longitudinal Stern-Gerlach
effect, which has been experimentally observed$^4$, appears to be
negligible in the present case.) 

The cumulative effect could be detected by
allowing the accelerated beam of neutrons to interfere with an
unaccelerated beam. Now in a neutron interferometry experiment the fringe
contrast is greatest when both interfering beams are in the same spin
state.  So, Wagh and Rakhecha have considered a cyclic evolution of the
spin state by letting the neutron go through another series of $2n$ cells
which amounts to a time reversal of the evolution that it underwent during
its journey through the first series of cells.  Then the dynamical
phase$^{5,6}$ acquired, which can in principle be observed in neutron
interferometry, depends on the momentum gained by the neutron as it passed
through the first $2n$ cells. However, since the phase shift is due to the
change in canonical momentum only, we cannot ambiguously conclude that its
detection is a verification of the new force.

A direct, unambiguous way of observing $\bf f$ is from the deviation of
the neutron beam due to the $y-$component of its kinetic momentum which it
acquires$^7$ as it passes through a sequence of these cells. This
component would be zero in the absence of the new force $\bf f$ because of
the translational symmetry in this direction. In the presence of it, $v_y$
is negative in each of the cells, and therefore the beam deviates
increasingly towards the negative $y-$ direction as it passes through the
cells. The detection of this deviation would constitute definitive
evidence of $\bf f$. Another way of observing $\bf f$
is to have neutrons incident with spin in the
$x-$direction and measure the time of flight through a large number of
cells, which would be modified by the change in the $x-$component of
velocity due to $\bf f$.

The observation $\bf f$ would provide direct evidence of the 
hidden momentum. If that is the only objective, it can be achieved
more simply by passing the 
neutron into a region of uniform $\bf E$-field, with no $\bf B$, 
such as the space between two capacitor plates, and measuring the 
velocity change due to the change in te hidden momentum ${\bf \mu}\times
{\bf E}$. For example, if $\bf E$ is in the 
$z-$ direction and the neutron spin is polarized along the x-
direction then, according to (2), the change of velocity is in the $y-$ 
direction equal 
to $\mu E\over m$. But the earlier proposal has the advantage 
that it would detect $\bf f$, which has never been observed 
before. Also, Yang-Mills fields are used for describing the weak 
and strong interactions which are short range, whereas, as 
mentioned, the magnetic dipole sees the electromagnetic field as a 
long range $SU(2)$ Yang-Mills field$^{1,2}$. So, observing $\bf f$ 
would demonstrate this important non abelian gauge field 
interaction, in addition to revealing a hitherto hidden aspect of the 
neutron or more generally a magnetic dipole.

I thank Apoorva Wagh for useful discussions.

\vskip .5cm
\noindent
{\it A shorter version of this article, but with two figures, appeared in
Nature, {\bf 387,}
558-559 (5 June 1997).}

\end{document}